\documentclass[osajnl,twocolumn,showpacs,superscriptaddress,10pt]{revtex4-1}

\usepackage{amsmath,amssymb,graphicx}

\begin{document}
\DeclareGraphicsExtensions{.eps,.pdf,.jpg,.bmp}
\title{Novel, Retroreflective, Magneto-Optical Trap Near a Surface}
\author{Jonathan Trossman}\email{Corresponding Author: \\ JonathanTrossman2014@u.northwestern.edu}
\affiliation{Department of Physics and Astronomy, Northwestern University, Evanston, Illinois 60209}

\author{Zigeng Liu}
\affiliation{Department of Physics and Astronomy, Northwestern University, Evanston, Illinois 60209}

\author{Ming-Feng Tu}
\affiliation{Department of Physics and Astronomy, Northwestern University, Evanston, Illinois 60209}

\author{Selim Shariar}
\affiliation{Department of Physics and Astronomy,
Department of Electrical and Computer Engineering, Northwestern University, Evanston, Illinois 60209}

\author{Brian Odom}
\affiliation{Department of Physics and Astronomy, Northwestern University, Evanston, Illinois 60209}

\author{J.B. Ketterson}\email{j-ketterson@northwestern.edu}
\affiliation{Department of Physics and Astronomy,
Department of Electrical and Computer Engineering, Northwestern University, Evanston, Illinois 60209}

\begin{abstract}
We report on a novel Magneto-Optical Trap (MOT) geometry involving the retroreflection of one of the six MOT beams in order to create an atom cloud close to the surface of a prism which does not have optical access along one axis.  A MOT of Rb$^{85}$ with $\sim 4 \times 10^7$ atoms can be created 700 $\mu$m from the surface.  The MOT lies close to the minimum of an evanescent Gravito-Optical Surface Trap (GOST) allowing for transfer into the GOST with potentially minimal losses.
\end{abstract}

\maketitle



The study of optical lattices has been an area of great interest in recent years \cite{greiner2002quantum}\cite{spielman2007mott}\cite{stoferle2004transition}\cite{sherson2010single}\cite{jordens2008mott}.  Surface traps are a prospective method for surpassing the diffraction limit for the lattice constants \cite{gerritsma2007lattice}\cite{gonzalez2015subwavelength}.  Here we will describe a technique for trapping cold Rubidium atoms close to a surface.  In the simple case we are considering, a boundary is generated by the evanescent field created by total internal reflection of a blue detuned laser beam, which with gravity creates a Gravito-Optical Surface Trap (GOST).  The first step to trapping in this potential is to create a MOT with a set of states that are well matched to those of a GOST.  In practice, this means creating the MOT close to the minimum of the GOST potential.

GOST traps have been studied in detail in the past, both on flat surfaces \cite{rychtarik2004crossover}\cite{aminoff1993cesium}\cite{hammes2000optical}\cite{soding1995gravitational}\cite{voigt2001elastic} and at the perimeter of optical fibers \cite{le2004atom} \cite{vetsch2010optical}.  For the planar case, the wavevector of the transmitted light is a function of wavelength, $\lambda$, index of refraction, $n$, and the angle of incidence, $\theta$, as given by

\begin{equation} \label{Evanescent Wavevector}
k = \sqrt{\left(\frac{2 \pi}{\lambda}\right)^2 - \left(\frac{2 \pi n sin(\theta)}{\lambda}\right)^2}
\end{equation}

Beyond the angle of total internal reflection this wavevector is imaginary, which gives rise to the exponential decay of the evanescent wave.  The amplitude of the field on the vacuum side of the boundary is given by the Fresnel coefficients.  Here we restrict to TM polarization;  with $E_i$ the field amplitude on the prism side of the boundary and $E_0$ is the field amplitude on the vacuum side of the boundary, the ratio given by

\begin{equation} \label{FresnellCoef}
\frac{E_0}{E_i} = \frac{2n cos(\theta)}{n^2cos(\theta) + \sqrt{n^2 - sin^2(\theta)}}
\end{equation}

The potential from a light source which is not resonant can be derived from 2nd order perturbation theory and is proportional to the light intensity and inversely proportional to the detuning of the frequency of light

\begin{equation} \label{Dipole Potential}
U_{dip}(r) = \frac{3\pi c^2 \Gamma}{2 \omega_0^3}\left(\frac{ I(r)}{\delta}\right)
\end{equation} 

Where $\Gamma$ is the spontaneous decay rate, $\omega_0$ is the resonant transition frequency, and $\delta$ is the difference between $\omega_0$ and the laser frequency. Combining equations \ref{FresnellCoef} and \ref{Dipole Potential} gives us the potential for the GOST as 

\begin{equation} \label{GOST_pot}
	U_{tot}(z) = \frac{3\pi c^2 \Gamma}{2 \omega_0^3}\left(\frac{ |E_0|^2}{\delta}\right) e^{-2|k|z} + mgz
\end{equation}

The distance from the potential minimum to the surface is dependent on the wavevector, $k$, the intensity of the evanescent wave, and the mass of the atoms.


Our MOT is configured for Rb$^{85}$ and has the geometry shown in Fig. 1.  A BK-7 right angle prism with a leg length of 30 mm and a width of 15 mm was cut from a 30 mm width unit supplied by Edmund Optics.  It is mounted on a specialized moveable prism holder in the vacuum chamber.

Two of the MOT beams are in the plane of the prism's hypotenuse and are adjusted so as to be close to the surface while simultaneously minimizing the scattering from the side of the prism.  These beams are then reflected back to make counter-propagating circularly-polarized beam pairs at the prism surface.  The third MOT beam is directed perpendicular to the plane of the hypotenuse of the prism.  This beam is then reflected via total internal reflection from the two legs of the prism after which it exits through the hypotenuse.  Each total internal reflection flips the handedness of the polarization, but since there are two reflections this has no effect.  The prism is not anti-reflection coated and therefore there is a small amount of reflection when the light enters the prism and this leads to a small change in the polarization.  This is compensated in the initial polarization as well as possible and does not inhibit the formation of a MOT.  Our prism geometry gives us some benefits over geometries such as that used in \cite{voigt2001elastic} due to the fact we can access the full range of angles for counterpropogating evanescent wave generating laser beams.

The triangular sides of the prism are not polished and therefore block some of the light.  This is minimized as much as possible but still has some effect as seen in the absorption images such as Fig.2.  Scattering from these sides leads to a false atomic signal as well as Fresnel diffraction of the probe beam used for absorption imaging.  The roughness of the cut side does not seem to be any more of a problem than the unpolished side.

\begin{figure} [htb] \label{fig:RetroMOTSch}
\centering
\includegraphics[scale=.3]{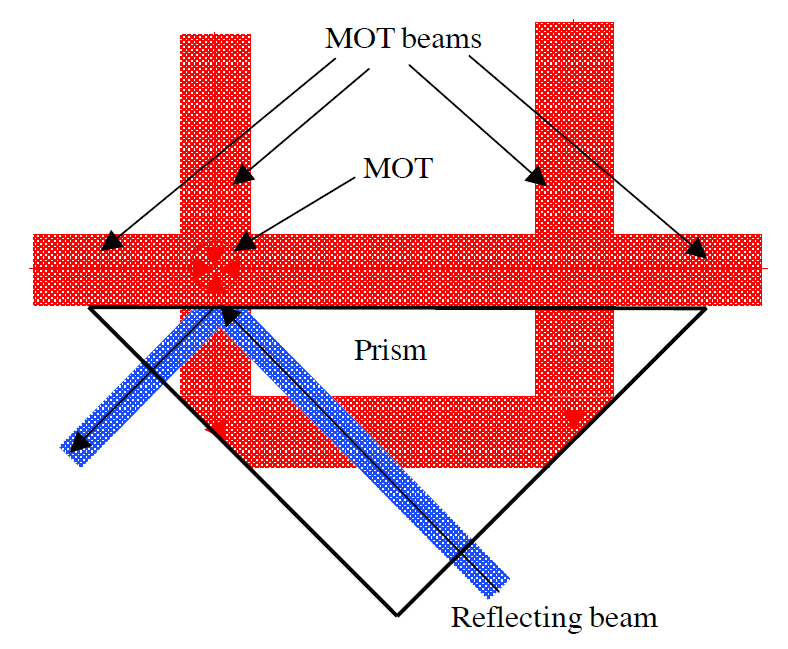}
\caption{Schematic of the retroreflection MOT geometry (the 3rd pair of MOT trapping beams is perpendicular to the page).  The incoming vertical beam is mostly $\sigma^+$.  The prism is not antireflection coated so the small amount of reflection causes a small change in the polarization.  This is corrected for with the $\lambda/4$ waveplate that polarizes this MOT beam.  Each retroreflection reverses the handedness of the polarization leaving the final beam's polarization very similar to the initial polarization.}
\end{figure}

This geometry allows us to position our MOT close to the surface as shown in Fig. 2, which is an absorption image of the MOT taken 100 $\mu$s after release of the cloud.  The absorption image is taken parallel to the prism surface.  We see that the highest density point of the MOT is $\sim$ 750 $\mu$m from the prism surface.  The elliptical shape of the MOT is possibly explained by a slight misalignment of the MOT beams, but also by real radiation pressure associated with light scattered from the vertical light beams by surface roughness on the prism.

\begin{figure} [htb]
\centering
\includegraphics[scale=.48]{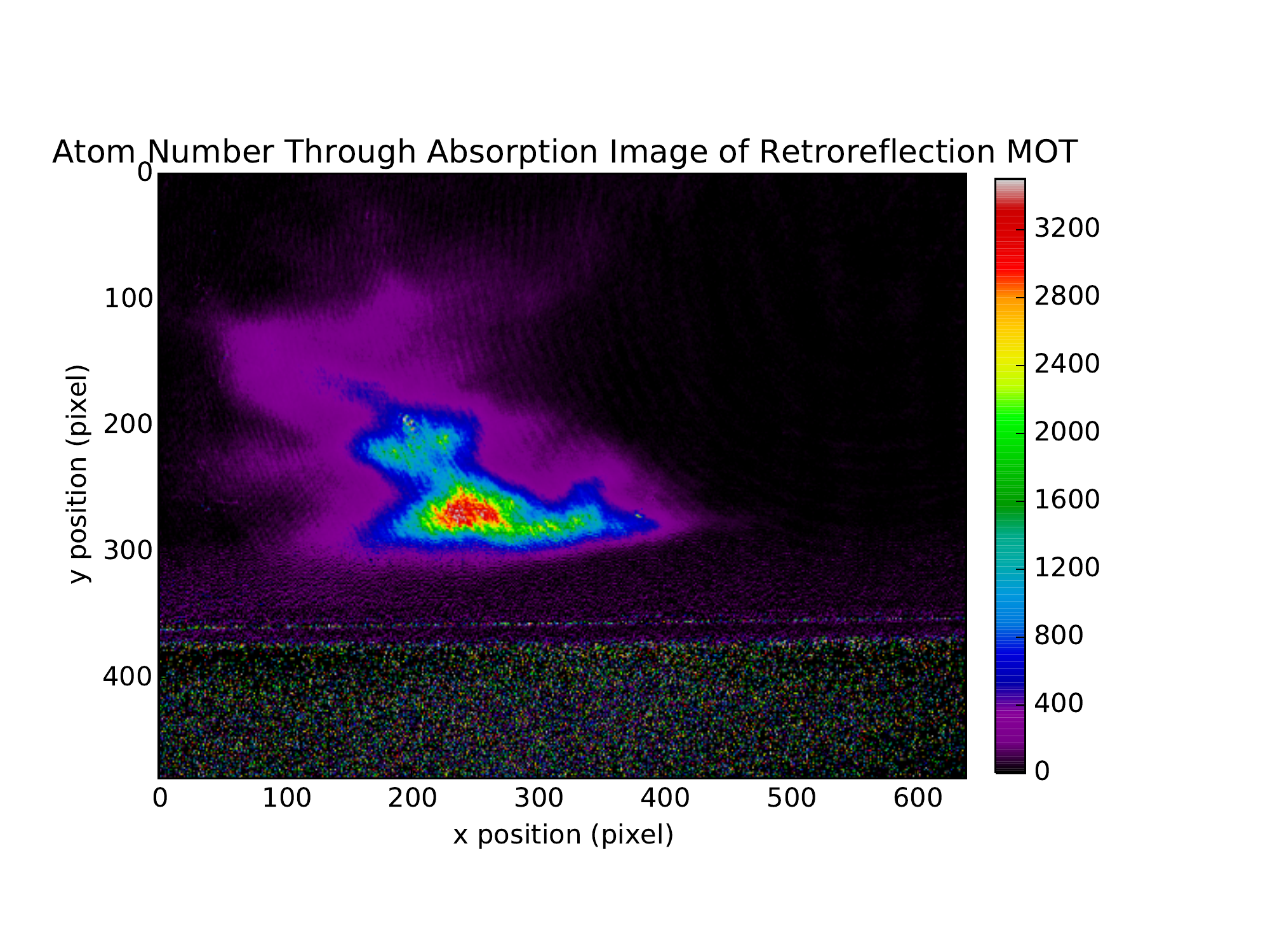}\label{fig:MOToverSurface}
\caption{Absorption image of the Rb$^{85}$ MOT above the surface averaged over 100 images and after 100 $\mu$s time of flight.  Pixels correspond to 7.5 $\mu$m.  The number of atoms is $\sim$ 4 $\times$ 10$^7$ corresponding to an average density of $\sim$ 1 $\times$ 10$^{10}$ atoms/cm$^3$. The absorption imaging beam is scattered by the side of the prism.  This leads to two effects.  First, there is a false signal of atoms where the surface blocks the beam.  Second, there is Fresnel diffraction of the probe beam leading to ripples near the surface.}

\end{figure}


Our total internally reflected laser beam is incident at 45$^o$, has 200 mW, and is 1 GHz blue detuned from the Rb$^{85}$ D2 line F=3 $\rightarrow$ F'=4.  Using Eq. \ref{GOST_pot}, it follows that the potential minimum will be 100 nm from the surface.  To observe this, we release the MOT and wait for 3000 $\mu$s.  The MOT will drop and expand, which is demonstrated in Fig. 3.  When we release the MOT in the presence of the total internal reflected beam, we see that the evanescent wave causes a deformation in the expansion of the cold atom cloud by reflecting the hottest atoms.

\begin{figure*} [htb]
\centering
\includegraphics[scale=.41]{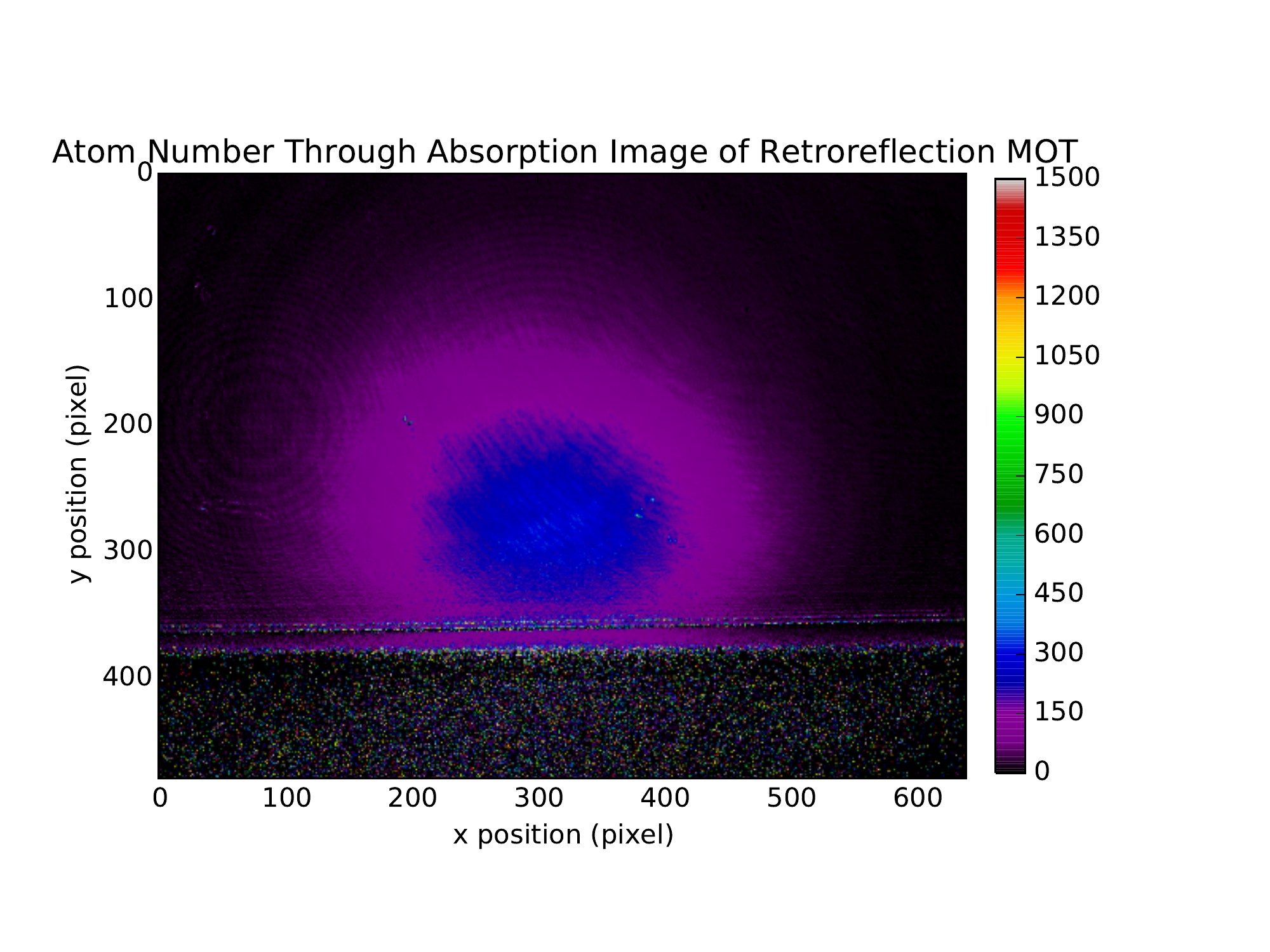}\label{fig:3000MOToverSurface}
\quad
\includegraphics[scale=.41]{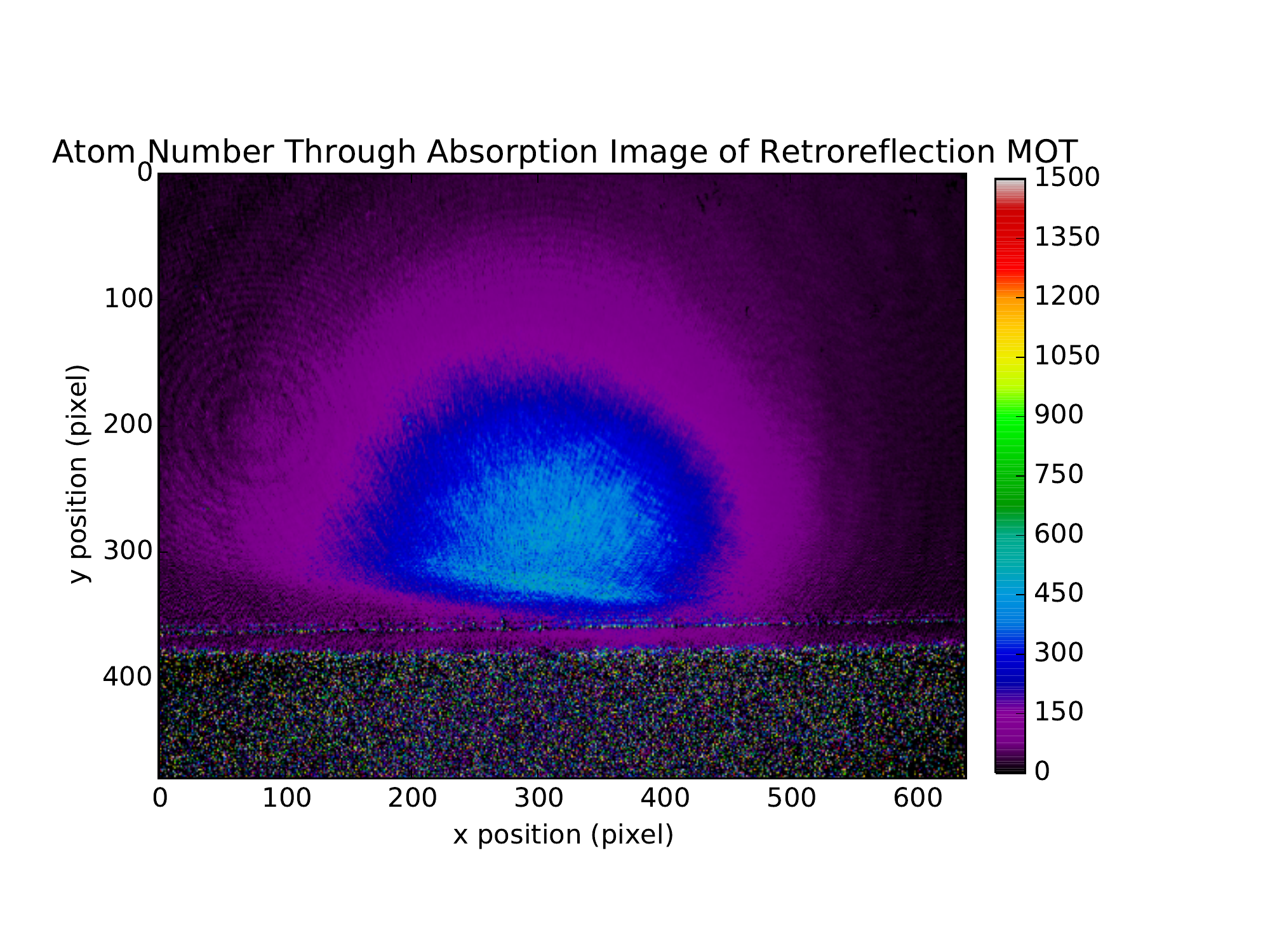}\label{fig:GOSTMOT}
\caption{The figure on the left is an absorption image of the Rb$^{85}$ cloud after a 3000 $\mu$s time of flight expansion.  We see the expected pseudo-momentum gaussian distribution.  The figure on the right is the same situation, but including the evanescent wave comprised of a 200 mW beam 1 GHz detuned at an incident angle of 45$^o$.  In this case we observe a deformation of the atomic cloud as the atoms approach the surface.}
\end{figure*}


A number of questions remain that can be examined with this retroreflection trap.  First, we notice that qualitatively the number of atoms appears to increase for a MOT close to an evanescent wave.  For future investigation, we will confirm that this is the case and investigate other effects an evanescent wave may have on the loading of a MOT.  Second, we will use the GOST to make a stable trap using surface Sisyphus cooling \cite{hammes2003evanescent} and to eventually create small lattice constant periodic potentials in the evanescent wave.


This work was supported through the NSF IGERT program under grant DGE-0801685 and through Northwestern's MRSEC program under NSF grant DMR-1121262.

\bibliographystyle{unsrt}
\bibliography{cite}

\begin{thebibliography}{10}

\bibitem{greiner2002quantum}
Markus Greiner, Olaf Mandel, Tilman Esslinger, Theodor~W H{\"a}nsch, and
  Immanuel Bloch.
\newblock Quantum phase transition from a superfluid to a mott insulator in a
  gas of ultracold atoms.
\newblock {\em Nature}, 415(6867):39--44, 2002.

\bibitem{spielman2007mott}
IB~Spielman, WD~Phillips, and JV~Porto.
\newblock Mott-insulator transition in a two-dimensional atomic bose gas.
\newblock {\em Physical review letters}, 98(8):080--404, 2007.

\bibitem{stoferle2004transition}
Thilo St{\"o}ferle, Henning Moritz, Christian Schori, Michael K{\"o}hl, and
  Tilman Esslinger.
\newblock Transition from a strongly interacting 1d superfluid to a mott
  insulator.
\newblock {\em Physical review letters}, 92(13):130403, 2004.

\bibitem{sherson2010single}
Jacob~F Sherson, Christof Weitenberg, Manuel Endres, Marc Cheneau, Immanuel
  Bloch, and Stefan Kuhr.
\newblock Single-atom-resolved fluorescence imaging of an atomic mott
  insulator.
\newblock {\em Nature}, 467(7311):68--72, 2010.

\bibitem{jordens2008mott}
Robert J{\"o}rdens, Niels Strohmaier, Kenneth G{\"u}nter, Henning Moritz, and
  Tilman Esslinger.
\newblock A mott insulator of fermionic atoms in an optical lattice.
\newblock {\em Nature}, 455(7210):204--207, 2008.

\bibitem{gerritsma2007lattice}
R~Gerritsma, S~Whitlock, T~Fernholz, H~Schlatter, JA~Luigjes, J-U Thiele,
  JB~Goedkoop, and RJC Spreeuw.
\newblock Lattice of microtraps for ultracold atoms based on patterned magnetic
  films.
\newblock {\em Physical Review A}, 76(3):033408, 2007.

\bibitem{gonzalez2015subwavelength}
A~Gonz{\'a}lez-Tudela, C-L Hung, DE~Chang, JI~Cirac, and HJ~Kimble.
\newblock Subwavelength vacuum lattices and atom--atom interactions in
  two-dimensional photonic crystals.
\newblock {\em Nature Photonics}, 9(5):320--325, 2015.

\bibitem{rychtarik2004crossover}
D~Rychtarik, B~Engeser, M~Hammes, H-C N{\"a}gerl, and R~Grimm.
\newblock Crossover to 2d in a double-evanescent wave trap.
\newblock In {\em Journal de Physique IV (Proceedings)}, volume 116, pages
  241--245. EDP sciences, 2004.

\bibitem{aminoff1993cesium}
CG~Aminoff, AM~Steane, P~Bouyer, P~Desbiolles, J~Dalibard, and
  C~Cohen-Tannoudji.
\newblock Cesium atoms bouncing in a stable gravitational cavity.
\newblock {\em Physical review letters}, 71(19):3083, 1993.

\bibitem{hammes2000optical}
M~Hammes, D~Rychtarik, V~Druzhinina, U~Moslener, I~Manek-H{\"o}nninger, and
  R~Grimm.
\newblock Optical and evaporative cooling of caesium atoms in the
  gravito-optical surface trap.
\newblock {\em Journal of Modern Optics}, 47(14-15):2755--2767, 2000.

\bibitem{soding1995gravitational}
J~S{\"o}ding, R~Grimm, and Yu~B Ovchinnikov.
\newblock Gravitational laser trap for atoms with evanescent-wave cooling.
\newblock {\em Optics communications}, 119(5):652--662, 1995.

\bibitem{le2004atom}
Fam Le~Kien, VI~Balykin, and K~Hakuta.
\newblock Atom trap and waveguide using a two-color evanescent light field
  around a subwavelength-diameter optical fiber.
\newblock {\em Physical Review A}, 70(6):063403, 2004.

\bibitem{vetsch2010optical}
E~Vetsch, D~Reitz, G~Sagu{\'e}, R~Schmidt, ST~Dawkins, and A~Rauschenbeutel.
\newblock Optical interface created by laser-cooled atoms trapped in the
  evanescent field surrounding an optical nanofiber.
\newblock {\em Physical review letters}, 104(20):203603, 2010.

\bibitem{hammes2003evanescent}
M~Hammes, D~Rychtarik, B~Engeser, H-C N{\"a}gerl, and R~Grimm.
\newblock Evanescent-wave trapping and evaporative cooling of an atomic gas at
  the crossover to two dimensions.
\newblock {\em Physical review letters}, 90(17):173001, 2003.

\end{thebibliography}

\end{document}